%% file: main.tex
\documentclass[conference]{IEEEtran}

\usepackage{xspace}

\newcommand{\projectName}{\emph{SENDIM}\xspace}

\usepackage{balance}  
\usepackage{graphicx} 
\usepackage{times}    
\usepackage{url}      
\usepackage[utf8]{inputenc}
\usepackage[usenames, dvipsnames]{color}
\usepackage{balance}
\usepackage{moreverb}
\usepackage{amsmath}
\usepackage[utf8]{inputenc}
\usepackage{url}

\usepackage{listings}
\usepackage{color}
\usepackage{textcomp}
\usepackage{algpseudocode}
\usepackage{algorithm}
\usepackage{lmodern}
\usepackage{enumerate}
\usepackage{varwidth}
\usepackage{xcolor}
\definecolor{mygreen0}{rgb}{0, 0.75, 0}

\definecolor{myred1}{rgb}{1,0,0}
\definecolor{mygreen1}{rgb}{0, 1, 0}
\definecolor{myblue0}{rgb}{0, 0, 1}

\definecolor{myred2}{rgb}{1,0.5,0.5}
\definecolor{mygreen2}{rgb}{0.5, 1, 0.5}
\definecolor{myblue2}{rgb}{0.5, 0.5, 1}

\definecolor{mygreen}{rgb}{0, 0.25, 0}
\definecolor{myblue}{rgb}{0, 0, 0.75}
\definecolor{myred0}{rgb}{0.5,0,0}

\definecolor{listinggray}{gray}{0.98}
\definecolor{lbcolor}{rgb}{0.98,0.98,0.98}
\begin{document}
\date{}
\bibliographystyle{IEEEtran}
\title{\LARGE \bf \projectName for Incremental Development of Cloud Networks\vspace{-1em}}

\author{{\bf Pradeeban Kathiravelu} \\
{\small INESC-ID Lisboa / Instituto Superior Técnico} \\
{\small Universidade de Lisboa, Portugal}\\
{\small pradeeban.kathiravelu@tecnico.ulisboa.pt}\\
\and 
{\bf Lu{\'\i}s Veiga} \\
{\small INESC-ID Lisboa / Instituto Superior Técnico} \\
{\small Universidade de Lisboa, Portugal}\\
{\small luis.veiga@inesc-id.pt}}
\maketitle

\subsection*{Abstract}
Due to the limited and varying availability of cheap infrastructure and resources, cloud network systems and applications are tested in simulation and emulation environments prior to physical deployments, at different stages of development. Configuration management tools manage deployments and migrations across different cloud platforms, mitigating tedious system administration efforts. However, currently a cloud networking simulation cannot be migrated as an emulation, or vice versa, without rewriting and manually re-deploying the simulated application. This paper presents \projectName\footnote{Sendim is a northeastern Portuguese town close to the Spanish border, where the rare Mirandese language is spoken.}, a \textbf{S}imulation, \textbf{E}mulation, a\textbf{N}d \textbf{D}eployment \textbf{I}ntegration \textbf{M}iddleware for cloud networks. As an orchestration platform for incrementally building Software-Defined Cloud Networks (SDCN), \projectName manages the development and deployment of algorithms and architectures the entire length from visualization, simulation, emulation, to physical deployments. Hence, \projectName optimizes the evaluation of cloud networks. 



\balance

\input{Introduction}

\input{related_work}	

\input{Architecture}

\input{Implementation}
	
\input{Evaluation}

\input{conclusion}



\bibliography{references}

\end{document}

%% file: Introduction.tex
\section{Introduction}
\label{sec:intro}

Cloud networks and architectures are tested over simulation, emulation, and physical test environments at different phases of implementation, as the required accuracy and precision in the evaluations differ, as the development progresses. Many of the network algorithms and design choices are tested for their functionality and efficiency in a simulation environment, prior to porting them to a physical deployment. While some parameters can be effectively estimated by a simulator, an emulator or a physical test environment will reveal more insights, as a simulator itself cannot provide an end-to-end guarantee on the testing of the system for accuracy and completion. 

As most of the publicly available networking simulators and emulators are developed independently from each other, they often consist of incompatible APIs and different development language and syntax. Learning and using them requires a considerable investment of time. Porting simulations to emulations or physical environments of varying system properties requires extensive code changes to the user algorithm or even a rewrite of majority of the code, with a manual redeployment over the environments. 

Configuration management tools such as Chef~\cite{spinellis2012don}, CFEngine~\cite{burgess1995cfengine}, and Puppet~\cite{turnbull2008pulling} manage the configurations of cloud-scale deployments. They automate the configuration and reconfiguration of the servers and virtual machines (VMs) in the cloud, eliminating the requirement to manually change or configure the servers, or write automation scripts. As they target the system administrators and cover mostly the deployment stage, configuring and deploying cloud applications for the early stage developments and quality assurance are still lacking. 

Network emulators such as Mininet~\cite{lantz2010network} are well-integrated into cloud and data center networks through the Software-Defined Networking (SDN) controllers, and provide a more realistic resemblance to the physical deployments. EmuSim is an integrated cloud simulation and emulation environment~\cite{calheiros2013emusim}, offering evaluation at different granularity and accuracy. However, unlike Mininet, EmuSim emulations cannot seamlessly be ported into enterprise cloud networks, and cannot be used to evaluate the production-ready algorithms in a physical deployment. A complete middleware platform should be developed offering an integrated management architecture and deployment process for the SDCN applications, from visualization, simulation, emulation, to physical deployments, leveraging the global view of the network available to the SDCN controllers. 

An integrated framework should enable incremental development and deployment of applications, as well as migration between different physical deployments. The configuration management approach can be extended to generalize the application logic such that a simulation can easily be migrated to an emulation environment, and to a physical deployment. While approaches leveraging design patterns such as service-oriented architecture (SOA)~\cite{perrey2003service} are used in managing multiple different components and products~\cite{manolescu2005service}, they do not target the existing SDCN applications. 

This paper presents \projectName, an integrated cloud management architecture for coordinating heterogeneous SDCN systems and components, extending the principles of SDN. \projectName offers an orchestration process for building, analyzing, and migrating efficient cloud network architectures with different topologies, design dimensions, and deployment environments. \projectName abstracts configurations and logic away from the application to provide seamless migration across multiple environments and different visualizations such as simulations and emulations. \projectName provides \textbf{\textit{Software-Defined Deployment}}, offering programmable deployments. Extending and leveraging SDCN for management of cloud deployments, \projectName focuses on minimizing the time to deliver of cloud applications.

In the upcoming sections, we will further analyze the proposed \projectName orchestration middleware platform for SDCN. We will continue to discuss the preliminary background information on the network simulators and emulators, while discussing SDN and cloud orchestration platforms, in Section 2. Section 3 discusses the design and architecture of the \projectName platform. Section 4 elaborates the implementation details of \projectName. Section 5 depicts the ongoing experimental evaluations on the prototype implementations, with the results discussed.  Finally, Section 6 will drive us to the conclusion of this research discussing its current state and the possible future enhancements.


%% file: related_work.tex
\section{Background and Related Work}
\label{sec:related_work}
\subsection{Network Emulators and Simulators}
NIST Net is a network emulator that statistically emulates an entire network in a single hop as a specialized router~\cite{carson2003nist}. Mininet emulates a network in a single machine~\cite{lantz2010network}, where Maxinet distributes Mininet emulation to a computer cluster~\cite{wette2014maxinet}.Open vSwitch is a virtual switch developed for hardware virtualization environments~\cite{pfaff2009extending}, and also is used by Mininet for emulating the switches. Moreover, Mininet emulations closely resemble physical networks, by leveraging the network virtualization offered by SDN~\cite{lantz2010network}. Servers emulated by Mininet can also emulate the execution of processes and applications installed in the host machine, and hence can be configured to offer an emulation of an entire cloud system inside a single computer.

Simulators are used when the considered system is too complex to emulate within the given time with the available resources, or in earlier stages of development when simulation is adequate. NS-3 provides a distributed simulation of networks~\cite{carneiro2010ns}, which can also be introduced into a live network, offering emulation capabilities to some extend~\cite{fall2005ns}. Frameworks with capability to both simulate and emulate the cloud networks, such as NS-3 and EmuSim, fail short in offering an enterprise production-quality solution, and they do not offer a seamless migration to physical cloud network deployments from simulations or emulations.

\subsection{Software-Defined Networking and Systems}

Cloud-scale networks tend to be large and have complex requirements, including privacy, security, service-level agreements (SLA), and business policies~\cite{kandukuri2009cloud}. Existing software-defined systems are developed independently, and they still lack a central orchestrator or a middlebox controller for all the physical and logical middlebox components that compose these systems, to be able to interoperate. These systems must be extended and leveraged to operate in a coordinated manner for a programmable enterprise cloud network. While there are previous works on applying SDCN for a finer grain control of cloud networks~\cite{benson2011cloudnaas}, further research is necessary to centrally orchestrate the computing, networking, and storage resources of the cloud-ready infrastructure. 

SDN controllers facilitate virtualization of networks, and enable programming the networks with an easy migration from emulated environments to physical environments, as network emulators can emulate cloud networks that resemble real network deployments. Migration from simulations to emulations is not straight-forward, and often requires designing the emulations and simulations separately. Leveraging the paradigm of SDN, simulators and emulators should be optimized to function in a more unified manner. However, modifying their APIs for an interoperability with each other or integration with an SDN controller require considerable effort due to their incompatible APIs and designs. Extending the SDN controllers, an approach and architecture should be devised for an integrated SDCN simulator-emulator to be able to have either simulations or emulations, based on the requirements.

\subsection{Cloud Orchestration}
Enterprises use multiple deployment environments during different stages of product development. Integrating simulation, emulation, and deployments for the heterogeneous systems to mitigate the administration overhead has been suggested for sensor networks~\cite{girod2004system}. Docker is an open source platform that provides an orchestration framework to build and execute distributed systems across different infrastructures locally and over clouds~\cite{merkel2014docker}. 

OASIS TOSCA\footnote{http://docs.oasis-open.org/tosca/TOSCA/v1.0/os/TOSCA-v1.0-os.html} is a standard for cloud orchestration and topology. OSGi provides a modular architecture~\cite{alliance2003osgi}, which is exploited in previous works in distributed and pervasive systems~\cite{rellermeyer2007r,gu2004toward,lee2003enabling}. Consisting of a componentized architecture, features can be developed and deployed as OSGi bundles, enabling an incremental development and deployment strategy across emulated and physical networked systems. Leveraging OSGi modular architecture for their extensibility, OpenDaylight~\cite{medved2014opendaylight} and ONOS~\cite{berde2014onos} controllers cover many aspects of SDN, providing network orchestration and service management functionality.

In a software development and testing process for cloud and network systems, automation and management tools focus on either deployment aspects, or early development stages such as simulation. While configuration management tools automate the deployment, reducing the manual efforts of DevOps~\cite{spinellis2012don}, they focus only on product deployment aspects and do not manage the early design and development configuration efforts. A framework that covers the entire development process to increase the productivity by minimizing the installation hell imposed due to the migration across multiple platforms should be researched and implemented for networking applications.


%% file: Architecture.tex
\section{Solution Architecture}

\label{sec:arch}
\projectName is an integration middleware that optimizes the evaluation of cloud network algorithms and architectures and minimizes the development efforts in testing the early research ideas to deployment, by offering a unified interface for simulation, emulation, and physical deployment of the cloud applications. It extends the motivation behind the SDN paradigm to the development aspects of applications that are designed to be executed on large scale SDCN systems consisting of many data centers. Our previous work $xSDN$~\cite{kathiravelu2015expressive}, an expressive simulator for network flows, has been extended to provide the cloud network simulation capabilities as well as a domain specific language with an easy-to-learn syntax to \projectName. 

Designed as an orchestration middleware framework for incremental development of SDCN applications, \projectName further leverages the software-defined systems to offer a logically centralized cloud networking controller to the early stage developments of the cloud networks, without limiting itself to the physical deployments. With the global view of the infrastructure available to the orchestrator, reusable cloud systems are composed and managed effectively.

\subsection{Software-Defined Cloud Orchestration}
\begin{figure}[ht]
	\begin{center}
			\resizebox{\columnwidth}{!}{
				\includegraphics[width=\textwidth]{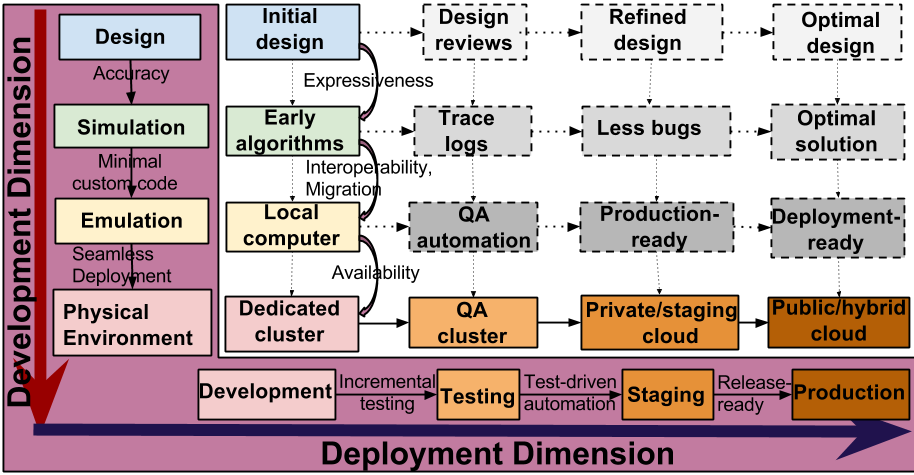}
			}
		\end{center}
		\caption{Software-Defined Deployments}
		\label{fig:depl}
	\end{figure}

\projectName offers a software-defined cloud orchestration, an integrated software engineering process for SDCN, that builds up on the interoperability offered by the software-defined systems, offering programmable deployments improving the testing and quality assurance of the developments since early stage, managing the entire development process from design to production. As shown by Figure~\ref{fig:depl}, software engineering process progresses in two dimensions: development dimension and deployment dimension. Development dimension consists of incrementally developing the application in different realizations, from design, simulation, emulation, to virtual and physical environments. The application is deployed in multiple environments such as development, testing, staging, to production environments.

In the development dimension, transition from higher to lower realizations imposes different challenges at each level. Simulators should provide accuracy in expressing the chosen design in simulation. Porting from simulation to emulation should have minimal custom code developed. A seamless deployment is mandatory when migrating from emulation to physical deployment. While simulations are suitable for a few application scenarios, emulations or physical test deployments are mandatory for other. \projectName leverages this for practical use case scenarios such as network load balancing and usage throttling algorithms, in building a system that scales itself across multiple realizations, with minimal human intervention.

Deployment dimension is test-driven in the enterprises. Development environment to testing environment should offer incremental testing. Test-driven automation~\cite{winkler2009test} enables smooth transition to staging environment. Staging is an environment identical to production, where release-ready applications are tested and migrated to production after final verifications. Multiple physical deployment environments exist based on the enterprise requirements. Development environment usually is a dedicated developer cluster. Testing environment consists of a QA cluster. Applications are often tested in a private or staging cloud, before moving to the production, such as a public or hybrid cloud.

In the deployment dimension, design is improved with revisions, to a refined and widely optimal design. Simulations are used to trace bugs with more logs and debugs in testing environment; with less bugs, it leads to an optimal solution. Emulations are migrated to testing environment with a QA automation. Emulations are production-ready in the staging environment phase, to deploy in production.

\projectName offers a software-defined deployment by separating the deployment logic from the details of the given environments, by building the applications as bundles and creating the revisions as changesets and updated bundles, that can be deployed in the application hosted in the respective environment. By decoupling the deployed application from the details of the individual development and deployment environments and realizations, \projectName automates and manages the entire matrix composed of these two dimensions. Hence, product deployments and updates are orchestrated from a central location more efficiently with less manual efforts.

\subsubsection{Anatomy of a \projectName Application}
Existing network simulators and emulators do not separate the application logic such as load balancing, application scheduling, and policy control in SDN, from the descriptions of the system that is simulated or emulated such as the properties of the data center and parameters of the network. Hence, the potential reuse of the application logic across simulation and emulation environments is prevented. In order to let applications to be migrated from simulation to emulation, the application to be simulated/emulated using \projectName is separated into the core application logic, the system to be simulated or emulated, and the information of deployment environment. Following a component-based software engineering approach~\cite{heydarnoori2008deploying}, development of \projectName applications focuses on separation of concerns (SoC).

The application logic is defined as a bundle and deployed into the SDN controller, where the system is either simulated by the simulation sandbox of \projectName, emulated by an emulator connected to the controller, or executed on a real physical network orchestrated by the controller. The separated logic allows reuse of code across simulation, emulation, and physical environments, hence enabling a unified application that functions as simulation, emulation, as well as virtual or physical execution of the real application, based on the deployment descriptions. While the application is built and tested iteratively as the development progresses, changesets are deployed accordingly. 

\begin{figure}[ht]
	\begin{center}
			\resizebox{\columnwidth}{!}{
				\includegraphics[width=\textwidth]{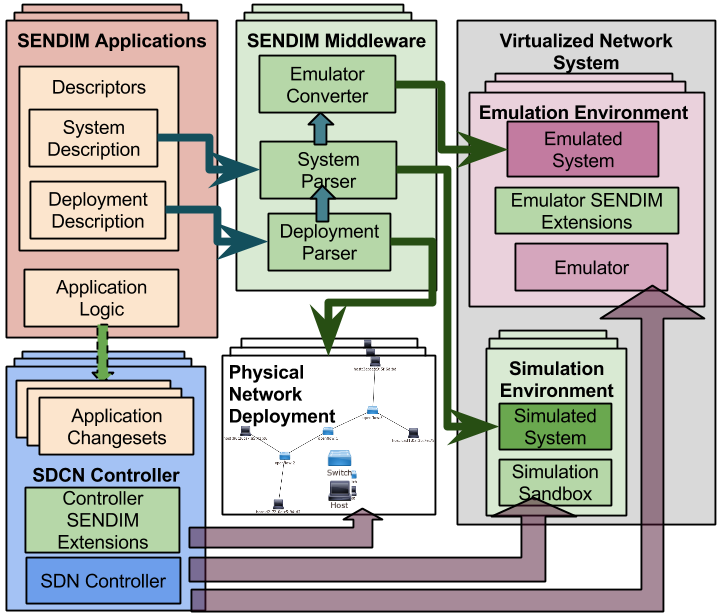}
			}
		\end{center}

		\caption{\projectName Middleware and Applications}
		\label{fig:ana}
	\end{figure}

Figure~\ref{fig:ana} shows the anatomy of a \projectName application, and how it is composed of the system and deployment descriptions and application logic. It also depicts the integration of the \projectName middleware into external hardware and software entities such as controllers, emulators, and physical networks. Deployment description is used in deciding whether the system is physical, or virtual - simulated or emulated, based on the system description. While the system description is native to the \projectName simulation sandbox, relevant \projectName emulator converters convert it into the emulation scripts for the emulator environments.

Multiple parallel applications can execute on the \projectName distributed middleware platform. Many emulation, simulation, and physical environments exist with different phases of developments. \projectName deploys the incremental developments of the application logic into the SDN controller as changesets. The effort on separating the application logic from the system and phase of the development is justified by the minimized effort on porting or migrating the application, and automated deployments with less manual interventions. 

\paragraph*{Expressiveness of \projectName} An XML-based domain specific language (DSL) is developed to represent and model highly complex software-defined cloud networks and systems, extending the DSL of xSDN~\cite{kathiravelu2015expressive}. Cloud systems and deployments are expressed in $descriptors$, the XML configuration files created abiding to the \projectName DSL, which define the syntax for expressing the application parameters and deployment details that can be separated from the applications. $Descriptors$ are either \textit{deployment descriptors} or \textit{system descriptors}. 

System properties such as nodes, links, and the properties of the nodes and links in the network and the dynamic properties that change frequently such as the flows that are executed on the system are defined in multiple configuration files known as the \textit{system descriptors}. A basic definition of a switch is shown below with the neighbors defined.
{\fontsize{9}{9}\selectfont				  
\begin{lstlisting}
  <node id="s1" type="switch">
            <link id="s2"></link>
            <link id="s5"></link>
    </node>
\end{lstlisting}
}
The granularity of the network flows can be set to flow, packet, or an intermediate chunk level where the flows are chunked into sub flows. A basic flow definition is shown below. 
{\fontsize{9}{9}\selectfont				  
\begin{lstlisting}
    <flow id="F1">
        <origin>h2</origin>
        <destination>h5</destination>
    </flow>
\end{lstlisting}
}

Similarly, which of the environment is used to deploy the algorithm is expressed in the \textit{deployment descriptor}. Frequent code changes are minimized as only changing the configuration files is sufficient to change the deployment or the dynamic properties of the system, and code changes are necessary only when the application logic is changed. Given below is a system definition for an emulation environment.
{\fontsize{9}{9}\selectfont				  
\begin{lstlisting}
    <system id="8">
        <type>emulation</type>
        <emulator>mininet</emulator>
        <home>~/mininet</home>
        <ctrlr>RemoteController</ctrlr>
        <controllerUrl>127.0.0.1</controllerUrl>
    </system>
\end{lstlisting}
}

\projectName provides expressiveness to simulation developments from the initial design, by letting the users compose systems with different topologies and nodes and flows in the network, by defining them in the \textit{system descriptors}, that can be generated from a graphical user interface manipulating the topology and the components. It offers interoperability and easy migration from simulations to emulations in a local computer of a developer. Based on the availability of the resources in a dedicated cluster, the emulated system can be migrated to a physical environment.

\subsubsection{\projectName Workflows}
User applications and system configuration processes are realized as workflows in \projectName. An application encompasses several phases of development, from simulation in development environment to physical deployment in a production environment. Algorithm~\ref{alg:start} presents the initializing of \projectName ecosystem as an extended SDCN.
\begin{algorithm}
{
		\fontsize{9}{9}\selectfont

	\caption{\projectName Middleware Execution}
	\label{alg:start}
	\begin{algorithmic}
		\Procedure{Initiate}{$emulators$, $controller$}
		\State $initInstances(controller)$
		\State $configurePhysicalNetworks()$
					\For {($emulator$ \textbf{in} $emulators$)}		
		\State $init(emulator)$
		\EndFor
		\State $initMiddleware()$
		\State $installExtensions(controller)$
							\For {($emulator$ \textbf{in} $emulators$)}		
		\State $installExtensions(emulator)$
		\EndFor

		\While{ $(true)$ }
		\State $executeControllerInstances()$
							\For {($emulator$ \textbf{in} $emulators$)}		
		\State $execute(emulator)$
		\EndFor
		\State $executeMiddleware()$
		\EndWhile
		\State $clearDistributedObjects()$
		\EndProcedure
	\end{algorithmic}

	}
\end{algorithm}

Initially, the distributed SDN controller instances are configured. Physical network deployments and emulator instances are then configured and initialized to be orchestrated by the controller. \projectName is initialized along with its simulation sandbox. \projectName installs the relevant extension packages into the controller and emulator instances to integrate them with \projectName. Once initialized, \projectName executes the controller and emulator instances till they are aborted by an interrupt. Once the \projectName middleware execution is terminated, the distributed packets of data and control flows are cleared from the cache.

Each application has an Initiator workflow, that deploys the initial version of the application into the controller and optionally simulate or emulate the system in the specified environment. It is also possible to deploy and execute the application into multiple environments at once using \projectName deployers. The entire application is deployed into the controller for the initial or first deployment of the algorithm. Subsequent iterative deployments only need to deploy the changesets. Pseudocode of the application initializer workflow is presented in Algorithm~\ref{alg:initiator}.

While physical deployments are identified and used to execute the workflows, emulations and simulations first need to construct the virtual networked systems before executing the network flows on them. In case of emulations, the system descriptors usually need to be converted to have the script for the respective emulator, whereas the simulations execute native inside \projectName, based on the system descriptor.


\begin{algorithm} [H]
	{
		\fontsize{9}{9}\selectfont

	\caption{\projectName Application Initialization}
		\label{alg:initiator}
		\begin{algorithmic}
					\Procedure{Deploy}{$deplDescriptor$, $sysDescriptor$, $controller$, $appBundles$}
		\State $deployment \gets $ $parse(deplDescriptor)$
		\State $deploy(controller, appBundles)$
		
				\If{$(deployment.env.isPhysical())$}
						\While{ $(TRUE)$ }
				\State $execute(deployment.env)$

		\EndWhile

				\ElsIf{$(deployment.env.isSimulation())$}
				\State $simulationSandbox \gets$ $construct($\\$deployment.env, sysDescriptor)$
						\While{ $(TRUE)$ }
	
				\State $execute(simulationSandbox)$
	
		\EndWhile

				\ElsIf{$(deployment.env.isEmulation())$}
				\State $emulator \gets deployment.env$
				\State $descriptor \gets convert(emulator, sysDescriptor)$
				\State $construct(emulator, descriptor)$
						\While{ $(TRUE)$ }

				\State $execute(emulator)$
		\EndWhile

		\EndIf
		
		\State $clearDistributedObjects()$
		\EndProcedure
		\end{algorithmic}

	}
\end{algorithm}

\subsubsection{Incremental Deployments and Migrations}
As development progresses, the application is tested, developed, and deployed incrementally and migrated across different platforms. Algorithm~\ref{alg:deploy} presents the subsequent deployments and migrations of the application which has already been deployed in the \projectName ecosystem.


\begin{algorithm}
{
		\fontsize{9}{9}\selectfont

	\caption{Iterative and Incremental Development}
	\label{alg:deploy}
	\begin{algorithmic}
	\Procedure{Upgrade}{$deplDescriptor$, $sysDescriptor$, $controller$, $appBundles$}
	\State $appID \gets appBundles.getID()$
\State $stopExperiment(appID)$
\State $clearDistributedObjects(appID)$
\State $clearState(appID)$
\State $changeSets \gets constructChangeSets(appBundles)$
\State \textbf{\textit{deploy}}($deplDescriptor$, $sysDescriptor$, $controller$, $appBundles$)
\EndProcedure
		\end{algorithmic}
	}

\end{algorithm}
The experiment is generally halted when the application is redeployed. If the application is the only experiment executing on the system, and the controller does not manage any other systems simultaneously, the controller can be simply restarted during this. However, in an orchestrated environment with multiple experiments and multiple networking systems, upgrades occur in a tenant-aware manner, pausing the execution of only the experiment by clearing the distributed objects and state of the application.

Changesets of the application are computed and built, since the last deployment. The application changesets are deployed into the environment consists of the controller and the physical or virtual network. A hot deployment~\cite{friese2004hot, irmert2007towards}, where applications are seamlessly upgraded in the controller, with minimal downtime, is possible if the application is updated, but not migrated to another environment. 

\subsection{\projectName Middleware Platform}
\begin{figure}[ht]
	\begin{center}
			\resizebox{\columnwidth}{!}{
				\includegraphics[width=\textwidth]{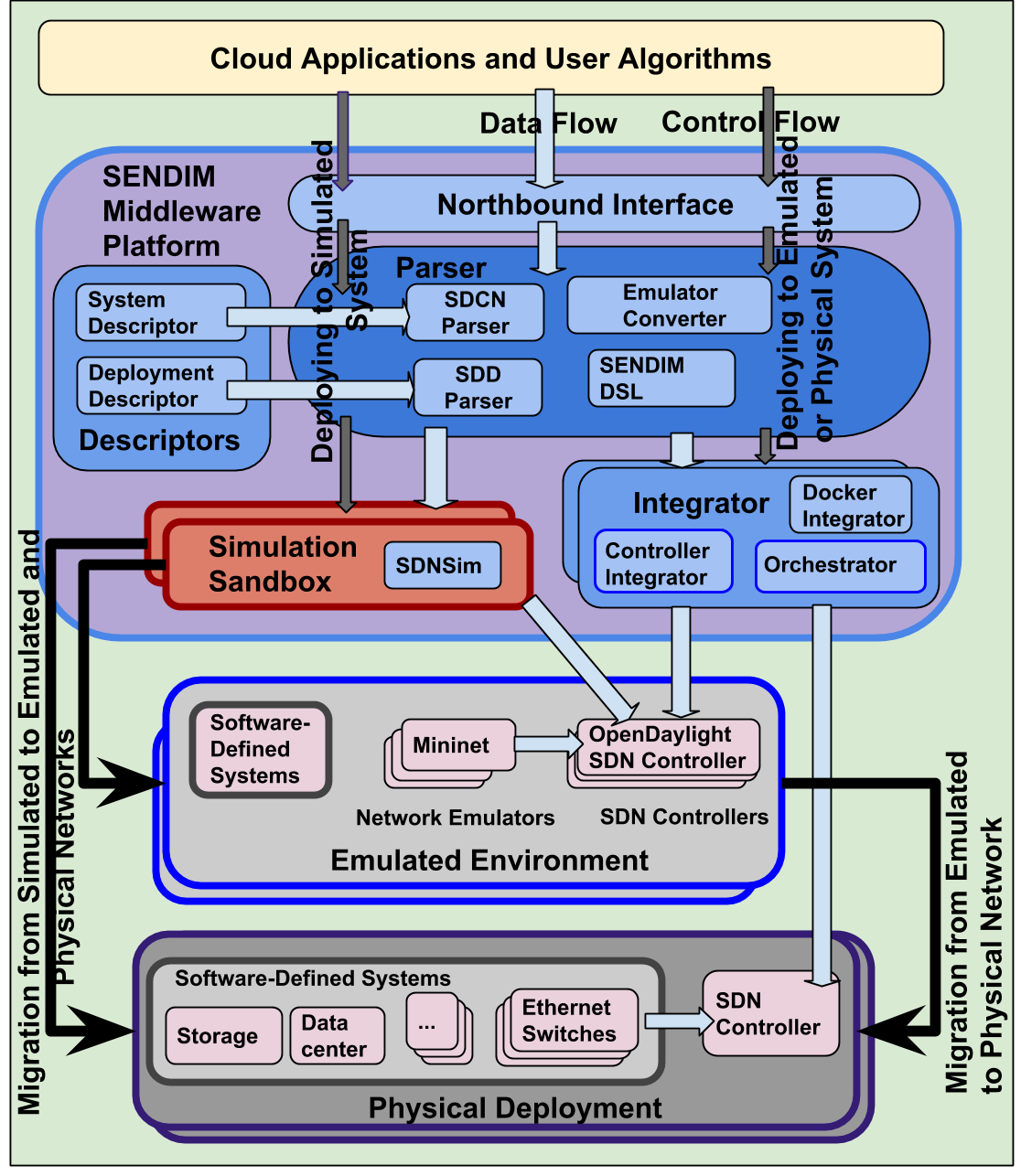}
			}
		\end{center}
		\caption{\projectName Architecture and Deployments}
		\label{fig:core}
		
	\end{figure}

The \projectName architecture can be realized as a distributed Java middleware platform to execute on multiple computer nodes, exploiting the in-memory data grids (IMDG)~\cite{marchioni2012infinispan}. OSGi component model is leveraged for a modular architecture. Figure~\ref{fig:core} provides a higher level deployment view, depicting the core components of the system. It depicts how the user algorithms and applications are deployed into either the simulation or emulation environment or deployed in physical networks, along with the cross-platform deployment migrations.

Adhering to the modular architecture, the components of \projectName are designed to be componentized and built as OSGi bundles, such that they can be deployed individually and executed independently. Hence, dependencies are defined for each bundles such that they will be satisfied when the bundles are deployed and installed independently and incrementally over the OSGi frameworks~\cite{irmert2007towards} or controllers such as OpenDaylight and ONOS which are built on top of the Apache Karaf OSGi framework.

\projectName northbound API is designed to be consistent with the existing network and cloud simulators such as CloudSim~\cite{calheiros2011cloudsim}, on top of which the user applications are executed. By adhering to a commonly used simulation interface, learning curve to the system is minimized in designing new simulations and migrating the existing ones to \projectName. The data and control flows pass through the northbound interface and \projectName parser to the simulation sandbox or controller integrator which forwards the data to the SDN controller. 

The \textit{SDCN parser} parses the \textit{system descriptors} to compose the software-defined systems for the simulation, where the \textit{SDD parser} parses the \textit{deployment descriptor} for software-defined deployments. The user applications are deployed and executed on the environment defined in a \textit{deployment descriptor} file, abiding to the deployment policies. While the SDCN is expressed in the system descriptor files following the \projectName DSL, the descriptors are automatically converted into the respective script for the emulators by the relevant Converter implementations. The relevant emulator is triggered with the network constructed and flows initialized, when the emulation environment is given as the execution environment.

The simulation sandbox provides simulation for software-defined networks and systems. While emulators and converters are connected through Converters and Integrators, simulation is handled in the $SDNSim$ simulation sandbox of \projectName. $SDNSim$ exploits the expressiveness and compactness of $xSDN$~\cite{kathiravelu2015expressive} network flow simulator, as it extends it into a simulator for SDN systems. The architecture of $SDNSim$ is compatible with the emulators, functionalities separated from the system descriptions. Currently existing cloud network simulators consist of considerable inconsistencies among the APIs and incompatibilities with the network emulator APIs, preventing them from being used as the simulator, without major code developments.

Controller integrator is an API for an integration with SDN controllers, together with which $SDNSim$ provides a software-defined deployment. Orchestrator provides extension to the SDN controller to connect with and orchestrate software-defined systems. Docker integrator wraps the application bundles to automate docker deployments.

Multiple algorithms can be simulated in parallel using a distributed simulation architecture~\cite{kathiravelu2014adaptive,kathiravelu2014concurrent}, with a physically distributed orchestrator. Creating a distributed in-memory data grid cluster, distributed and parallel simulations can execute in a single computer, a computer cluster, or a cloud environment. \projectName distributed executions can be made adaptive and elastic, following the distribution and scalability approaches proposed in the previous work for in-memory data grid applications~\cite{kathiravelu2014adaptive}. Thus more complex applications can be executed with minimal resources in a shorter time, consuming less energy and cost.

\projectName implementation includes the development of the middleware platform as well as the applications and scripts that are developed on top of \projectName as use cases. \projectName middleware includes the core \projectName including the orchestrator and \projectName core that provides a northbound API for the application developers. Moreover, it also includes extensions deployed into the controller and emulators for connecting them to \projectName middleware core by extending their east/westbound APIs. The extension points are for providing the emulation parameters for the emulator or for invoking the REST APIs to inject the \projectName control flow. Application changesets are deployed exploiting the northbound API of the controller, where the integration of the controller to the emulators, simulators, and physical deployments leverages the southbound API.
					
\subsubsection{SDN Simulation}
$SDNSim$ is the core bundle of \projectName, a compact and generic SDN simulator. Developed as an OSGi bundle, $SDNSim$ integrates with the OSGi-based SDN controllers seamlessly. $SDNSim$ mimics the Mininet API on defining the network composed of switches and hosts. It builds topologies by expressing the network system using XML configuration files. Flows can be recreated from the log files. In order to maintain compatibility with emulation environments, the simulation engine is kept light-weight with an easily extensible API to model more complex networks. The network is built when the simulation is started, and algorithms are simulated. Sample network topologies and deployment patterns are designed such that they can easily be extended to represent more complicated real-network.

Network emulators such as Mininet usually only emulate the network. Though application processes can be executed from the simulated hosts and switches, these processes are not emulated. Rather, they are the real applications executed from the host environment. Simulation environment just provides the measures and properties of the hosts such as the processing power and installed VM. Hence, they do not execute real processes for the simulated applications.

\subsubsection{Controller Integration}
Controller integrator is a southbound \projectName API, that is implemented for each of the SDN controllers depending on the controller's northbound API. With SDN controllers in place, virtual emulated switches can easily be replaced by Ethernet switches connected to the cloud networks. By connecting to an SDN controller, an emulated or physical system consisting of a software-defined network can be used to execute the logic from the algorithms.

Applications such as multiple implementations of network throttling~\cite{day2010network} and network load balancing~\cite{coughlin2002network} based on user profiles are designed and deployed into the controller such that the emulations and simulations execute on top of the controller can leverage the algorithm. While emulated networks do an actual execution of the algorithms over virtual switches and hosts, $SDNSim$ simulates the execution by representing the networked system using relevant Java objects, rather than actually executing the application across switches and hosts. Hence, a network of thousands of nodes can be simulated and visualized from the controller in a similar manner to the emulated networks, offering an appearance similar to the emulated network. Thus, much larger networks can be visualized and simulated, which is impossible in an emulated network due to the limitations in available resources. 

Due to its modular componentized architecture and abundantly available extension bundles, OpenDaylight controller is used as the default SDN controller of \projectName. Algorithms can be built on top of \projectName, and dynamically deployed in OpenDaylight as OSGi bundles, without changing their code. OpenDaylight Virtual Tenant Network (VTN) provides a multi-tenant virtual network over the controller, and hence is used for creating a tenant-aware orchestration environment for \projectName for multi-user parallel experiments.

Changes in user applications are built in the orchestrator instance and deployed into the relevant environment. Remote file copying and transferring is used to securely update the respective physical deployment. Scripts with remote calls are used to invoke or restart the remote execution, and to invoke the automation tools and packet installers. As OSGi framework manages the versioning of the plugins, when updated bundles are deployed, \projectName will upgrade itself and/or the deployed user algorithms to the new version upon the restart of the framework, without removing or altering the existing bundles. Hence, the deployed system could be reverted to a previous version, by removing the malfunctioning changesets or later versions of the bundles from the repository. This is exploited as a checkpointing feature~\cite{koo1987checkpointing} for \projectName, where development changesets are tagged and deployed before a major update or a milestone, enabling a quick revert to the previously known stable working state, should critical test cases fail.

\subsubsection{Emulator Integration}
Emulators are integrated to \projectName through the converters and emulator extensions. Converter converts the system descriptors into the scripts for the respective emulator. Emulator extensions enable remote invocations of the respective emulators to remotely execute, control, pause, and exit the executing instances. Multiple emulator instances execute from different deployment environments, leveraging the single distributed controller environment in a tenant-aware manner, hence enabling an easy migration across multiple platforms, from simulations to emulations and vice versa. 

Mininet has been leveraged as the default emulator. As emulations consume more resources than simulations, it is essential that the simulated platform should be able to be emulated in the \projectName system with the given resources. With minimal overheads during the transition and execution in addition to the base emulation, the overhead imposed by the middleware is designed to be negligible once the emulator is integrated. If the resources are available, a network system can be preconfigured to resemble the emulated system, such that a seamless migration is possible between the chosen emulation to physical deployment environment, without any modification to the controller.

%% file: Implementation.tex
\section{Implementation}
\label{sec:Implementation}
A prototype of \projectName platform\footnote{\url{https://sourceforge.net/projects/s2dn/.}} has been implemented based on the design. Oracle Java 1.8.0 is used as the development language, along with Apache Maven 3.1.1 to manage and build the bundles and execute the scripts. An in-memory data grid platform, Infinispan 7.2.0.Final~\cite{marchioni2012infinispan} has been exploited as the distributed shared knowledge base, providing a distributed execution and a physically distributed orchestrator with a logically centralized view. OpenDaylight Lithium is used as the default controller of the \projectName platform. Apache Karaf 3.0.3 is used as the OSGi run time by \projectName, as it is exploited for the bundle management by OpenDaylight, providing remote access and a security framework, in addition. Mininet 2.2.1 has been used as the default emulator.

Open vSwitch OF13 switches are simulated by programmatically invoking the OpenDaylight Switch interfaces and representation, for the integration of \projectName simulation sandbox into the OpenDaylight controller. Model-Driven Service Abstraction Layer (MD-SAL) of OpenDaylight offers an extensible and seamless deployment and installation of additional features. The simulation capabilities of $SDNSim$ are incorporated by invoking MD-SAL. 

\projectName applications or algorithms can be deployed into OpenDaylight as separate bundles, and the deployed algorithms can be simulated or emulated by the systems including $SDNSim$. $SDNSim$ can operate as a stand-alone simulator as well as a module of \projectName. It can function as an OpenDaylight project, when it is deployed as an OSGi bundle into the Apache Karaf container of OpenDaylight. The SDN system is simulated in $SDNSim$ and plugged into OpenDaylight controller by invoking its southbound API. 

YANG~\cite{bjorklund2010yang}, a data modeling language for the network configuration protocol (NETCONF)~\cite{schonwalder2010network} is used in defining the higher-level northbound abstraction of \projectName applications that are deployed in or integrated to OpenDaylight, hence enabling the connection to the simulation of virtualized network by $SDNSim$, and further configuring the network based on the simulated application. Simulated or emulated systems can be viewed from the OpenDaylight graphical user interface (DLUX), including the simulated or virtual switches and servers, flows and their properties, and other properties that can be extended to be exposed through the DLUX interface.

The properties are stored as distributed maps in \projectName. More properties can be defined with the property names as the keys and property values as the values of the map. Topologies and flows defined in XML following the \projectName DSL are translated into the relevant emulator scripts. For the emulation with Mininet, the defined DSL is converted into Python scripts by the \projectName Converter implementation for Mininet, which emulates the network and flows by invoking the Mininet instance. 

\projectName consists of W3C DOM~\cite{wang2007space} parsers which parse system and deployment information. $SDNSim$ simulates the system and executes the algorithm on the simulated system, when the \textit{deployment descriptor} indicates a simulation environment as the deployment platform. In case of emulation and physical deployments, the user applications are executed on top of the cluster consisting of the physical network, or which hosts the emulator instances. The componentized modular architecture provided by OpenDaylight and \projectName platform enables dynamic updates. The updated bundles are built and deployed dynamically into the execution environments.

Converter and integrators are invoked programmatically in Java as part of the \projectName middleware. Application logic is built using Maven, and deployed into the controller as changesets using shell scripts. MD-SAL extensions are implemented for OpenDaylight controller to deploy the applications as OSGi bundles, and execute them over the simulated, emulated, or physical networking system.

%% file: Evaluation.tex
\section{Evaluation}
\label{sec:Eval}
\projectName prototype was evaluated for the cloud orchestration features in integrated simulated and emulated environments, on a computer with Intel\textregistered\ Core\texttrademark\ i7-4700MQ CPU @ 2.40GHz × 8 processor, 8 GB memory, and Ubuntu 14.04 LTS 64 bit operating system, and compared with related frameworks for features and effectiveness. Multiple (up to 6) identical computers in a cluster, were used in evaluating the migration of the applications across different environments, each providing different deployment environments for the systems, from development to production.

\subsection{Comparative Qualitative Assessment}
\begin{figure}[ht]
	\begin{center}
			\resizebox{\columnwidth}{!}{
				\includegraphics[width=\textwidth]{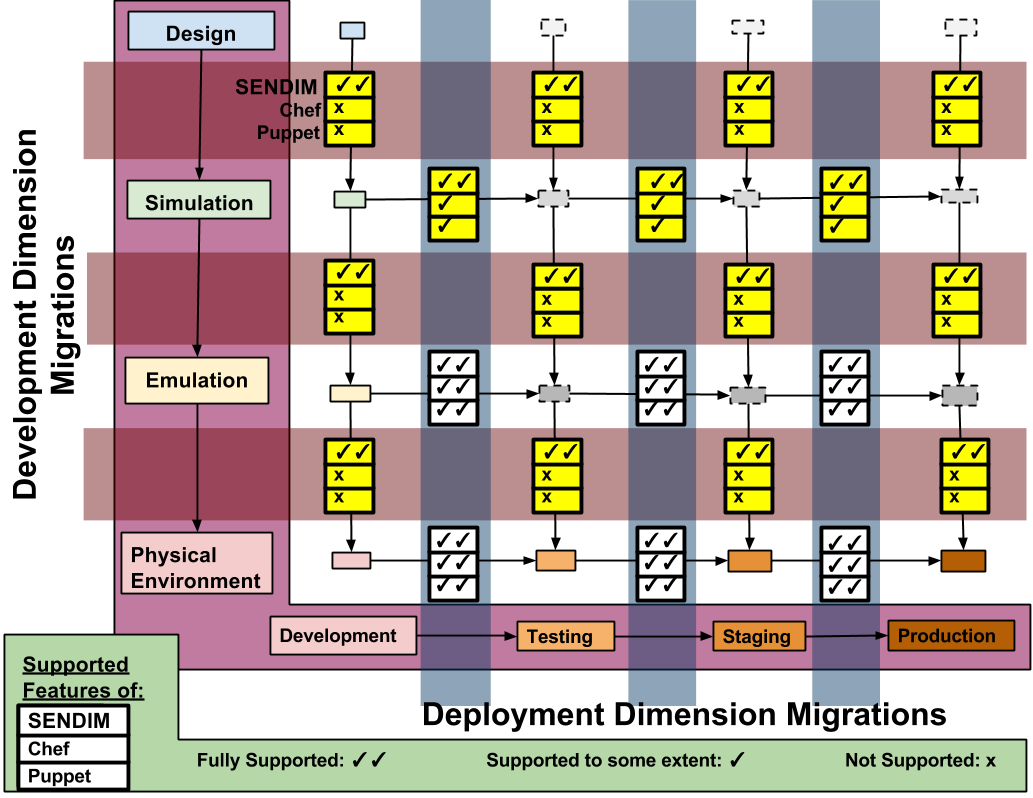}
			}

		\end{center}
		\caption{Comparison of Features}
		\label{fig:eval}
	\end{figure}

Configuration management systems offer automation for managing on-site and public cloud deployments with varying configurations, and orchestrate and manage multiple products hosted on them. Cloud orchestration and software-defined deployment functionalities offered by \projectName was compared with the popular configuration management systems, Puppet and Chef, to highlight further use cases and migrations covered by \projectName, as shown by Figure~\ref{fig:eval}. Transitions across both dimensions indicate, the migrations across different deployment environments for the same development dimension (in horizontal arrows), and migrations across different development dimensions in the same deployment environment (in vertical arrows). The vertical sets showed on transitions across both dimensions indicate the functionality offered by \projectName, Chef, and Puppet respectively, highlighting when \projectName is the best.

Configuration management systems function on migrating across different deployment environments of the same development dimension, providing a varying degree of support for each development dimension. For example, while emulations and physical deployments can effectively be handled and managed by the configuration management systems to migrate across different deployment environments, they are not efficient in handling simulations. Migration across different development dimensions is not achievable by the configuration management systems, as that is beyond the scope of these platforms. \projectName offers complete support for migration across both dimensions -- migrating across different deployment environments for the same development dimension, or migrating across different development dimensions over the same deployment environment. \projectName manages an entire matrix of development and deployment dimensions, with extension points to deployment systems such as Docker. Hence, it provides easy migration across more environments.

While both Puppet and Chef provide programmable infrastructure, they do not provide extensive support for software-defined infrastructure, leveraging the reusability and programmability offered by SDN and software-defined storage or cover the management across the development dimension which are offered by \projectName. Both \projectName and Puppet provide an easy-to-use declarative language as DSL, where Chef uses a Ruby-based DSL. 


Manual actions were reduced to 10\% in a typical SDCN application life cycle, with actions such as deploying the application, starting or restarting the controller, migrating the simulation scripts to emulation script, starting or restarting the emulator, incremental deployments of application bundle changesets, and starting or restarting the simulator are automated by Maven scripts and workflows executed through a single Java entry point of \projectName framework.

\subsection{Assessment of Incremental Development of Cloud} 
\projectName was then evaluated for its efficiency and performance in functioning as an orchestration middleware for the incremental development of SDCN. Time taken for multiple operations was measured and recorded. Small-scale experiments were repeated 6 times and the average time was considered, in order to mitigate the efforts of external factors influencing the execution time. OpenDaylight was used as the controller during the evaluation, where Mininet was used as the default emulator platform. Sparsely connected networks with random and uniform links across switches were constructed for the evaluation. Average degree, which is the average of the number of links per each node was maintained up to 100 during the experiments. End-to-end routing between two hosts and ping-all are used as the sample experiments.

\subsubsection{Seamless migration from emulation mode to simulation mode}
\begin{figure}[H]
	\begin{center}
			\resizebox{\columnwidth}{!}{
				\includegraphics[width=\textwidth]{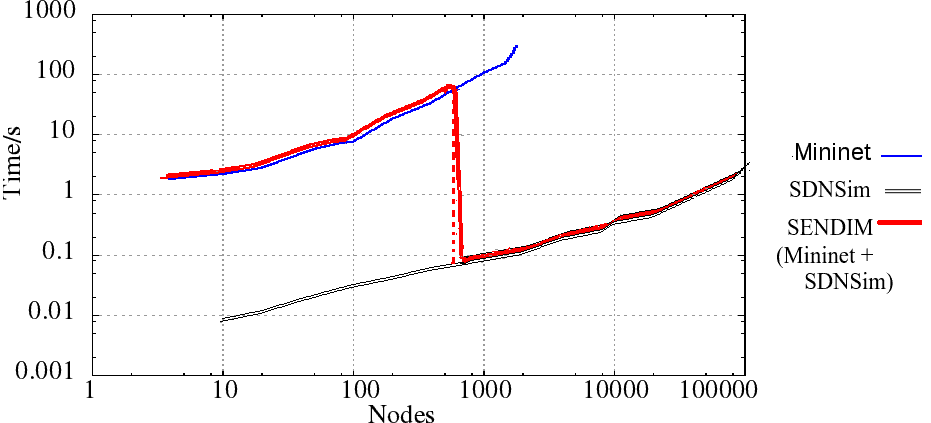}
			}
		\end{center}

		\caption{Network Construction with Mininet and \projectName SDNSim}
		\label{fig:mn}
	\end{figure}

A network was constructed with varying number of switches and hosts and with an average degree of 1, with Mininet and SDNSim. Figure~\ref{fig:mn} shows the time taken for Mininet to emulate the system, SDNSim to simulate the system, and \projectName configured with automated scripts to emulate or simulate adaptively based on the resources.

\projectName was configured to resort to SDNSim simulations when the emulations take more than 1 minute. This approach offers best of both worlds, as it offers the accuracy of emulations when resources are available, and simulate for larger networks with limited resources, transitioning seamlessly across both realizations. Moreover, this also indicates the potential of a seamless migration from simulations to emulations by either increasing the computing resources, or by reducing the simulated system size.

While SDNSim was able to construct the system with nodes and links within a few seconds even for large systems with up to 100,000 nodes, Mininet consumed up to 300 seconds even for a network of 1800 nodes and links. Hence, SDNSim showed a 100-fold performance compared to Mininet in visualizing the network. When increasing the network size further up to 1900, Mininet failed to create the links, though it successfully emulated the nodes. For network beyond the size of 8000, Mininet fails with an OSError when attempting to construct the nodes. When emulating a network of 1800 nodes with Mininet, Network Manager consumed as high as 99.6\% of CPU during the links construction phase, and up to 100\% of CPU consumption by the Open vSwitch daemon (ovs-vswitchd) was noticed during the final stages of emulation. However, simulation of the same network with $SDNSim$ consumed only up to 6\% of CPU.

Results as accurate as those produced by Mininet emulation were observed with SDNSim simulations for simple network tasks such as routing algorithms, cloud resource allocation, and network load balancing. This allows \projectName to resort to simulations to visualize large scale data centers in the early stages of development.


\subsubsection{Simulation to Emulation Migrations}
Mininet converter for \projectName has been evaluated for its efficiency in migrating the \projectName SDNSim simulations into Mininet emulations, with the efforts reduced in developing separate custom code for simulations and emulations. Complex networks with small-world datacenter topology~\cite{shin2011small} were modeled using the DSL of \projectName, and a routing was simulated across multiple hosts in shortest path. Based on the deployment descriptor, the solution was either simulated in \projectName SDNSim simulation sandbox, or automatically converted by \projectName into the relevant Python script and emulated using Mininet without manual coding or deploying. \projectName converter was used to migrate the simulations to emulations, where the system descriptions in XML were converted to Mininet scripts.

\begin{figure}[!ht]
	\begin{center}
			\resizebox{\columnwidth}{!}{
				\includegraphics[width=\textwidth]{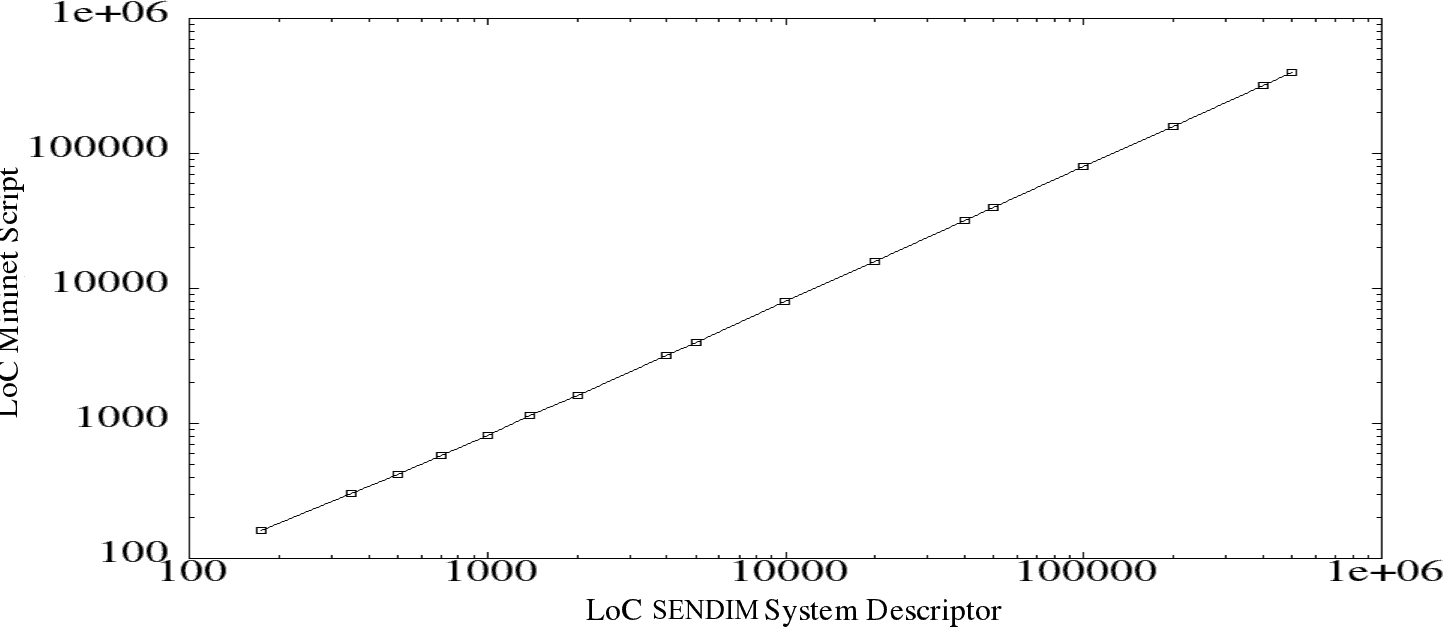}
			}
		\end{center}

		\caption{Performance of Code Conversion and Autogeneration}
		\label{fig:loc}
	\end{figure}

Simulation to emulation migration was experimented with varying number of nodes including switches and hosts and varying average degree. System descriptors expressed the complex systems which are parsed into the network as well as the flows and custom properties. Figure~\ref{fig:loc} indicates the lines of code (LoC) in the system descriptors of various \projectName systems that are developed, and the lines of autogenerated Python code, eliminating the requirement to re-write the code for Mininet emulator.

The converter showed linear transformation of code, scaling well for large XML files. Algorithms were deployed in the Karaf container of OpenDaylight, and not manually modified for the migration, though incremental deployments are possible. Hence, with a 25\% of additional lines from emulation code as shown by Figure~\ref{fig:loc}, both simulation and emulation are provided. This is estimated to save 37.5\% (as 100\% + 25\%, making 125\% effort for both simulation and emulation, saving (200 - 125) / 2 \% of effort) of the development efforts for each simulation and emulation, compared to the case where the simulation and emulation are done separately. A sample resource allocation scenario required 3900 LoC for an NS-3 simulation, and 4020 LoC for a Mininet emulation, where \projectName offered both simulation and emulation for 5000 LoC.

\begin{figure}[!ht]
	\begin{center}
			\resizebox{\columnwidth}{!}{
				\includegraphics[width=\textwidth]{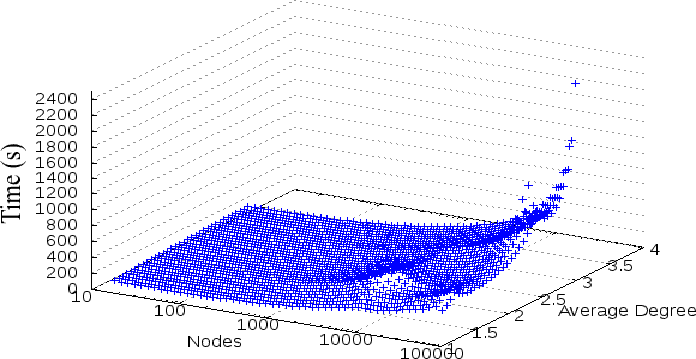}
			}
		\end{center}
		\caption{Migrating a Simulation to Emulation}
		\label{fig:mnc}
	\end{figure}

Figure~\ref{fig:mnc} shows the time taken to migrate an application from simulation to emulation, by converting the system descriptor scripts, initiating Mininet with the network consists of nodes, which are either switches or hosts, and links across those nodes as specified in the descriptor, and start the Mininet execution of simple routing across the nodes. Time taken for the migration increases with the number of nodes, as the converter has to read, parse, and convert the descriptor files into Mininet script, and start the execution. Links are expressed as neighbor nodes to each of the nodes, and hence fine-grain manipulation of flows is possible with accurate representation of real SDCN systems.

There was also an increase in time with the increase in the number of average degrees per node, which is visible more in larger networks with more nodes. As seen by the graph, \projectName was able to handle 10,000 nodes within a few seconds. Larger and complex networks of 100,000 nodes were successfully migrated, though there was a clearly above linear increase in the time taken. This was because of the memory and processing required for the parsing of large descriptor files which are as large as 26 MB, and writing Mininet scripts that are of up to 18 MB for 100,000 nodes with up to 100 links for each node.

Larger networks can be migrated by using a distributed cluster for the migration in a shorter time, though a single computer was good enough for the current data center-scale systems. Currently no existing framework is able to migrate a cloud-scale network from simulation to deployment without rewriting the code, where \projectName separates the logic and deploys the application logic in an SDN controller, orthogonal to the realizations such as simulation or emulation, and automate the deployments and migrations across multiple systems and realizations based on the deployment descriptor.

%% file: conclusion.tex
\section{Conclusion and Future Work}
\label{sec:conclusion}
\projectName minimizes manual deployments by automating deployments and migrations across different platforms and infrastructures. We foresee a shorter time to deliver as \projectName minimizes the development overhead caused by incompatible APIs, programming languages, and granularity among the simulators and emulators, and administrative overheads caused by the deployment migration. 

\projectName integrates and automates multiple phases of development process, by enabling seamless migration through loose coupling of algorithms from their deployment environments and development realizations such as simulations and emulations. The experimental evaluation elaborated the potential benefits of the middleware platform in increasing the productivity and efficiency of the SDCN application development cycle. A complete cloud platform can be orchestrated using the proposed architecture, providing an end-to-end guarantee on testing the user applications.

\projectName prototype should be extended to fully function as a complete middleware platform for generalizing the migrations by implementing controller integrator API for different SDN controllers, as well as the application deployment platforms such as Docker~\cite{merkel2014docker}. Implementation of extension points to enable easy migration across more deployment infrastructures and platforms, and a complete evaluation of the platform for more production scenarios and use cases are ongoing and future works.



%% file: main.bbl
\begin{thebibliography}{10}
\providecommand{\url}[1]{#1}
\csname url@samestyle\endcsname
\providecommand{\newblock}{\relax}
\providecommand{\bibinfo}[2]{#2}
\providecommand{\BIBentrySTDinterwordspacing}{\spaceskip=0pt\relax}
\providecommand{\BIBentryALTinterwordstretchfactor}{4}
\providecommand{\BIBentryALTinterwordspacing}{\spaceskip=\fontdimen2\font plus
\BIBentryALTinterwordstretchfactor\fontdimen3\font minus
  \fontdimen4\font\relax}
\providecommand{\BIBforeignlanguage}[2]{{%
\expandafter\ifx\csname l@#1\endcsname\relax
\typeout{** WARNING: IEEEtran.bst: No hyphenation pattern has been}%
\typeout{** loaded for the language `#1'. Using the pattern for}%
\typeout{** the default language instead.}%
\else
\language=\csname l@#1\endcsname
\fi
#2}}
\providecommand{\BIBdecl}{\relax}
\BIBdecl

\bibitem{spinellis2012don}
D.~Spinellis, ``Don't install software by hand,'' \emph{Software, IEEE},
  vol.~29, no.~4, pp. 86--87, 2012.

\bibitem{burgess1995cfengine}
M.~Burgess \emph{et~al.}, ``Cfengine: a site configuration engine,'' in
  \emph{in USENIX Computing systems, Vol}.\hskip 1em plus 0.5em minus
  0.4em\relax Citeseer, 1995.

\bibitem{turnbull2008pulling}
J.~Turnbull, \emph{Pulling strings with puppet: configuration management made
  easy}.\hskip 1em plus 0.5em minus 0.4em\relax Springer, 2008.

\bibitem{lantz2010network}
B.~Lantz, B.~Heller, and N.~McKeown, ``A network in a laptop: rapid prototyping
  for software-defined networks,'' in \emph{Proceedings of the 9th ACM SIGCOMM
  Workshop on Hot Topics in Networks}.\hskip 1em plus 0.5em minus 0.4em\relax
  ACM, 2010, p.~19.

\bibitem{calheiros2013emusim}
R.~N. Calheiros, M.~A. Netto, C.~A. De~Rose, and R.~Buyya, ``Emusim: an
  integrated emulation and simulation environment for modeling, evaluation, and
  validation of performance of cloud computing applications,'' \emph{Software:
  Practice and Experience}, vol.~43, no.~5, pp. 595--612, 2013.

\bibitem{perrey2003service}
R.~Perrey and M.~Lycett, ``Service-oriented architecture,'' in
  \emph{Applications and the Internet Workshops, 2003. Proceedings. 2003
  Symposium on}.\hskip 1em plus 0.5em minus 0.4em\relax IEEE, 2003, pp.
  116--119.

\bibitem{manolescu2005service}
D.~Manolescu and B.~Lublinsky, ``Service orchestration patterns: graduating
  from state of the practice to state of the art,'' in \emph{Companion to the
  20th annual ACM SIGPLAN conference on Object-oriented programming, systems,
  languages, and applications}.\hskip 1em plus 0.5em minus 0.4em\relax ACM,
  2005, pp. 148--149.

\bibitem{carson2003nist}
M.~Carson and D.~Santay, ``Nist net: a linux-based network emulation tool,''
  \emph{ACM SIGCOMM Computer Communication Review}, vol.~33, no.~3, pp.
  111--126, 2003.

\bibitem{wette2014maxinet}
P.~Wette, M.~Draxler, and A.~Schwabe, ``Maxinet: distributed emulation of
  software-defined networks,'' in \emph{Networking Conference, 2014
  IFIP}.\hskip 1em plus 0.5em minus 0.4em\relax IEEE, 2014, pp. 1--9.

\bibitem{pfaff2009extending}
B.~Pfaff, J.~Pettit, K.~Amidon, M.~Casado, T.~Koponen, and S.~Shenker,
  ``Extending networking into the virtualization layer.'' in \emph{Hotnets},
  2009.

\bibitem{carneiro2010ns}
G.~Carneiro, ``Ns-3: Network simulator 3,'' in \emph{UTM Lab Meeting April},
  vol.~20, 2010.

\bibitem{fall2005ns}
K.~Fall and K.~Varadhan, ``The ns manual (formerly ns notes and
  documentation).''

\bibitem{kandukuri2009cloud}
B.~R. Kandukuri, V.~R. Paturi, and A.~Rakshit, ``Cloud security issues,'' in
  \emph{Services Computing, 2009. SCC'09. IEEE International Conference
  on}.\hskip 1em plus 0.5em minus 0.4em\relax IEEE, 2009, pp. 517--520.

\bibitem{benson2011cloudnaas}
T.~Benson, A.~Akella, A.~Shaikh, and S.~Sahu, ``Cloudnaas: a cloud networking
  platform for enterprise applications,'' in \emph{Proceedings of the 2nd ACM
  Symposium on Cloud Computing}.\hskip 1em plus 0.5em minus 0.4em\relax ACM,
  2011, p.~8.

\bibitem{girod2004system}
L.~Girod, T.~Stathopoulos, N.~Ramanathan, J.~Elson, D.~Estrin, E.~Osterweil,
  and T.~Schoellhammer, ``A system for simulation, emulation, and deployment of
  heterogeneous sensor networks,'' in \emph{Proceedings of the 2nd
  international conference on Embedded networked sensor systems}.\hskip 1em
  plus 0.5em minus 0.4em\relax ACM, 2004, pp. 201--213.

\bibitem{merkel2014docker}
D.~Merkel, ``Docker: lightweight linux containers for consistent development
  and deployment,'' \emph{Linux Journal}, vol. 2014, no. 239, p.~2, 2014.

\bibitem{alliance2003osgi}
O.~Alliance, \emph{Osgi service platform, release 3}.\hskip 1em plus 0.5em
  minus 0.4em\relax IOS Press, Inc., 2003.

\bibitem{rellermeyer2007r}
J.~S. Rellermeyer, G.~Alonso, and T.~Roscoe, ``R-osgi: distributed applications
  through software modularization,'' in \emph{Proceedings of the
  ACM/IFIP/USENIX 2007 International Conference on Middleware}.\hskip 1em plus
  0.5em minus 0.4em\relax Springer-Verlag New York, Inc., 2007, pp. 1--20.

\bibitem{gu2004toward}
T.~Gu, H.~K. Pung, and D.~Q. Zhang, ``Toward an osgi-based infrastructure for
  context-aware applications,'' \emph{Pervasive Computing, IEEE}, vol.~3,
  no.~4, pp. 66--74, 2004.

\bibitem{lee2003enabling}
C.~Lee, D.~Nordstedt, and S.~Helal, ``Enabling smart spaces with osgi,''
  \emph{Pervasive Computing, IEEE}, vol.~2, no.~3, pp. 89--94, 2003.

\bibitem{medved2014opendaylight}
J.~Medved, R.~Varga, A.~Tkacik, and K.~Gray, ``Opendaylight: Towards a
  model-driven sdn controller architecture,'' in \emph{2014 IEEE 15th
  International Symposium on}.\hskip 1em plus 0.5em minus 0.4em\relax IEEE,
  2014, pp. 1--6.

\bibitem{berde2014onos}
P.~Berde, M.~Gerola, J.~Hart, Y.~Higuchi, M.~Kobayashi, T.~Koide, B.~Lantz,
  B.~O'Connor, P.~Radoslavov, W.~Snow \emph{et~al.}, ``Onos: towards an open,
  distributed sdn os,'' in \emph{Proceedings of the third workshop on Hot
  topics in software defined networking}.\hskip 1em plus 0.5em minus
  0.4em\relax ACM, 2014, pp. 1--6.

\bibitem{kathiravelu2015expressive}
P.~Kathiravelu and L.~Veiga, ``An expressive simulator for dynamic network
  flows,'' in \emph{Cloud Engineering (IC2E), 2015 IEEE International
  Conference on}.\hskip 1em plus 0.5em minus 0.4em\relax IEEE, 2015, pp.
  311--316.

\bibitem{winkler2009test}
D.~Winkler, S.~Biffl, and T.~{\"O}streicher, ``Test-driven automation--adopting
  test-first development to improve automation systems engineering processes,''
  in \emph{16th EuroSPI Conference}, 2009.

\bibitem{heydarnoori2008deploying}
A.~Heydarnoori, ``Deploying component--based applications: Tools and
  techniques,'' in \emph{Software Engineering Research, Management and
  Applications}.\hskip 1em plus 0.5em minus 0.4em\relax Springer, 2008, pp.
  29--42.

\bibitem{friese2004hot}
T.~Friese, M.~Smith, and B.~Freisleben, ``Hot service deployment in an ad hoc
  grid environment,'' in \emph{Proceedings of the 2nd international conference
  on Service oriented computing}.\hskip 1em plus 0.5em minus 0.4em\relax ACM,
  2004, pp. 75--83.

\bibitem{irmert2007towards}
F.~Irmert, M.~Meyerh{\"o}fer, and M.~Weiten, ``Towards runtime adaptation in a
  soa environment.'' \emph{RAM-SE}, vol.~7, pp. 17--26, 2007.

\bibitem{marchioni2012infinispan}
F.~Marchioni, \emph{Infinispan Data Grid Platform}.\hskip 1em plus 0.5em minus
  0.4em\relax Packt Publishing Ltd, 2012.

\bibitem{calheiros2011cloudsim}
R.~N. Calheiros, R.~Ranjan, A.~Beloglazov, C.~A. De~Rose, and R.~Buyya,
  ``Cloudsim: a toolkit for modeling and simulation of cloud computing
  environments and evaluation of resource provisioning algorithms,''
  \emph{Software: Practice and Experience}, vol.~41, no.~1, pp. 23--50, 2011.

\bibitem{kathiravelu2014adaptive}
P.~Kathiravelu and L.~Veiga, ``An adaptive distributed simulator for cloud and
  mapreduce algorithms and architectures,'' in \emph{Utility and Cloud
  Computing (UCC), 2014 IEEE/ACM 7th International Conference on}.\hskip 1em
  plus 0.5em minus 0.4em\relax IEEE, 2014, pp. 79--88.

\bibitem{kathiravelu2014concurrent}
------, ``Concurrent and distributed cloudsim simulations,'' in
  \emph{Modelling, Analysis \& Simulation of Computer and Telecommunication
  Systems (MASCOTS), 2014 IEEE 22nd International Symposium on}.\hskip 1em plus
  0.5em minus 0.4em\relax IEEE, 2014, pp. 490--493.

\bibitem{day2010network}
M.~S. Day, ``Network connection detection and throttling,'' Oct.~12 2010, uS
  Patent 7,814,542.

\bibitem{coughlin2002network}
C.~Coughlin, ``Network load balancing,'' Jun.~28 2002, uS Patent App.
  10/185,329.

\bibitem{koo1987checkpointing}
R.~Koo and S.~Toueg, ``Checkpointing and rollback-recovery for distributed
  systems,'' \emph{Software Engineering, IEEE Transactions on}, no.~1, pp.
  23--31, 1987.

\bibitem{bjorklund2010yang}
M.~Bjorklund, ``Yang-a data modeling language for the network configuration
  protocol (netconf),'' 2010.

\bibitem{schonwalder2010network}
J.~Sch{\"o}nw{\"a}lder, M.~Bj{\"o}rklund, and P.~Shafer, ``Network
  configuration management using netconf and yang,'' \emph{IEEE Communications
  Magazine}, vol.~48, no.~9, pp. 166--173, 2010.

\bibitem{wang2007space}
F.~Wang, J.~Li, and H.~Homayounfar, ``A space efficient xml dom parser,''
  \emph{Data \& Knowledge Engineering}, vol.~60, no.~1, pp. 185--207, 2007.

\bibitem{shin2011small}
J.-Y. Shin, B.~Wong, and E.~G. Sirer, ``Small-world datacenters,'' in
  \emph{Proceedings of the 2nd ACM Symposium on Cloud Computing}.\hskip 1em
  plus 0.5em minus 0.4em\relax ACM, 2011, p.~2.

\end{thebibliography}
